\documentclass[twocolumn,aps,amssymb,superscriptaddress]{revtex4}
\usepackage{graphicx}

\begin{document}
\title{On the peak in the far-infrared conductivity of strongly anisotropic cuprates}
\author{A. Pimenov}
\author{A. V. Pronin}
\altaffiliation[permanent address: ]{General Physics Institute of the Russian Acad. of
Sciences, 119991 Moscow, Russia}
\author{A. Loidl}
\affiliation{Experimentalphysik V, Elektronische Korrelationen und Magnetismus, Institut
f\"{u}r Physik, Universit\"{a}t Augsburg,
86135 Augsburg, Germany} %
\author{A. Tsukada}
\author{M. Naito}
\affiliation{NTT Basic Research Laboratories, 3-1, Morinosato-Wakamiya, Atsugi-shi, Kanagawa 243-0198, Japan}%

\date{\today}

\begin{abstract}
We investigate the far-infrared and submillimeter-wave conductivity of electron-doped
La$_{2-x}$Ce$_{x}$CuO$_{4}$ tilted $1^\circ$ off from the ab-plane. The effective
conductivity measured for this tilt angle reveals an intensive peak at finite frequency
($\nu \sim 50\,$cm$^{-1}$) due to a mixing of the in-plane and out-of-plane responses.
The peak disappears for the pure in-plane response and transforms to the Drude-like
contribution. Comparative analysis of the mixed and the in-plane contributions allows to
extract the c-axis conductivity which shows a Josephson plasma resonance at $\nu \simeq
11.7\,$cm$^{-1}$ in the superconducting state.
\end{abstract}

\maketitle

Far-infrared conductivity in the normally-conducting state of the high-$T_{\rm c}$
cuprates can be well characterized by the Drude-like response \cite{ir}. This response
leads to the levelling-off of the real part of the conductivity $\sigma^* =\sigma_1 +
i\sigma_2$ in the submillimeter frequency range \cite{ybco,ncco2}. In some cases
deviations from this behavior can be observed in the conductivity spectra as broad maxima
at far-infrared frequencies. One possible reason for such additional excitation can be a
charge localization induced by doping \cite{basov98}. More generally, disorder, such as
inhomogeneous oxygen distribution \cite{mcguire,puchkov,singley}, has been shown to lead
to a an additional maximum in conductivity \cite{atkinson}. Further possible effects
leading to a conductivity peak include charge density waves \cite{bernhard} or reduced
dimensionality in the static stripe phase \cite{dumm}.

In this paper we present another possible mechanism for finite-frequency peak in the
infrared conductivity of electron-doped cuprate La$_{2-x}$Ce$_{x}$CuO$_{4}$ (LCCO). The
same arguments will hold for strongly anisotropic, mostly two-dimensional metals in which
the out-of-plane response is nearly insulating. In the present case the peak arises due
to a small admixture of the c-axis response to the in-plane conductivity. Rotating the
polarization of the infrared radiation, the pure in-plane response can also be measured.
The in-plane conductivity shows no infrared peak and follows the predictions of the Drude
model. The comparative analysis of the spectra for different experimental geometries
allowed to extract conductivity and dielectric constant along the c-axis.

La$_{2-x}$Ce$_{x}$CuO$_{4}$ film with $x=0.075$ was deposited by molecular-beam epitaxy
\cite{naito,naito1} on transparent SrLaAlO$_{4}$ substrate $10\times10\times 0.5$ mm$^3$
in size. The thickness of the film was 140 nm and the film revealed a sharp transition
($\Delta T_{c} < 1$\,K) at 25\,K. Lower transition temperature compared to the
optimally-doped compound ($T_{\rm c}=30\,$K) \cite{naito1} and low Ce-concentration
indicate the underdoped regime in the film. The SrLaAlO$_{4}$ has been cut close to the
(001)K direction with the misfit angle of $\alpha=1^\circ \pm 0.1^\circ$. In this case
the film grows nearly c-axis oriented with the same off-axis tilt-angle.

In order to obtain the complex conductivity in the frequency range from
submillimeter-waves to far-infrared, two different experimental methods \cite{mgb2ir}
have been applied to the same film. For frequencies below 40\,cm$^{-1}$ the complex
conductivity of the film has been obtained by the submillimeter transmission spectroscopy
\cite{volkov}. At higher frequencies the sample reflectance has been measured using
standard far-infrared techniques and the conductivity was obtained via a Kramers-Kronig
analysis of the spectra.

The transmission experiments for frequencies $5$\,cm$^{-1}<\nu <40$\,cm$^{-1}$ were
carried out in a Mach-Zehnder interferometer arrangement \cite{volkov} which allows both,
the measurements of the transmittance and the phase shift of a film on a substrate. The
properties of the blank substrate were determined in a separate experiment. Utilizing the
Fresnel optical formulas for the complex transmission coefficient of the substrate-film
system, the absolute values of the complex conductivity $\sigma ^{*}=\sigma _{1}+i\sigma
_{2}$ were determined directly from the measured spectra. In the frequency range $40<\nu
<4000$\,cm$^{-1}$ reflectivity measurements were performed using a Bruker IFS-113v
Fourier-transform spectrometer. In addition, the reflectance for frequencies $5<\nu
<40$\,cm$^{-1}$ has been calculated from the complex conductivity data of the same
sample, obtained by the submillimeter transmission. This substantially improves the
quality of the subsequent Kramers-Kronig analysis of the reflectance and therefore the
reliability of the data at the low-frequency part of the infrared spectrum.

\begin{figure}[]
\centering
\includegraphics[width=7cm,clip]{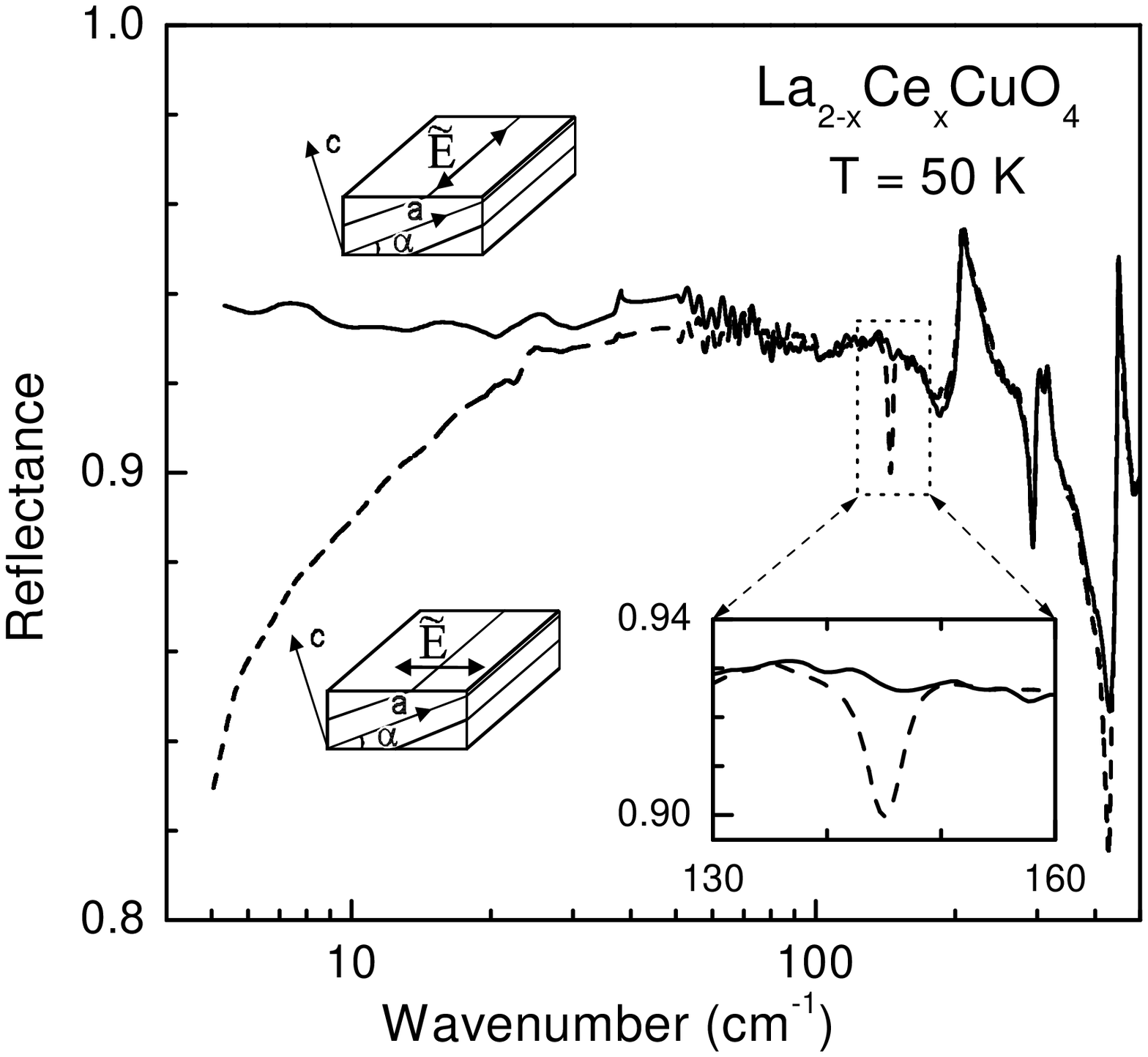}
\caption{Far-infrared reflectance of an underdoped LCCO film at different experimental
geometries at $T=50\,$K. Solid lines - pure in-plane response, dashed lines - tilted
geometry. The reflectance data above 40\,cm$^{-1}$ were measured directly and the spectra
below this frequency were calculated from the complex conductivity data obtained by the
transmittance technique. The inset shows the magnified spectra around a longitudinal
c-axis phonon.} \label{frefl}
\end{figure}

Due to the tilt-structure of the film, two main experimental geometries are possible.
Both cases are depicted in Fig. \ref{frefl}. In the first case (upper drawing) the
currents, induced by the incident radiation, are within the CuO$_2$ plains only.
Therefore, the pure in-plane response is obtained in this case. In the other experimental
geometry (lower drawing in Fig. \ref{frefl}) a small admixture of the c-axis response is
expected.

Figure \ref{frefl} shows the far-infrared reflectance of the LCCO film for both possible
experimental geometries. Sharp structures above 200 cm$^{-1}$ are attributed to the
phonons in the substrate. In that case no substantial difference is seen between the
reflectance in both geometries. In addition to these effects, sharp minima in reflectance
can be observed, which reveal a clear polarization dependence. One of such resonances is
shown in the inset. We interpret these resonances as LO phonons in LCCO. The most
important differences between the spectra of both geometries are seen below
50\,cm$^{-1}$: the spectra of the mixed geometry deviate significantly for decreasing
frequencies.


The reflectivity of a thin metallic film on a dielectric substrate can be obtained from
the Maxwell equations \cite{heavens}:
\begin{equation}
r=\frac{r_{0f}+r_{fs}\exp ({4\pi }in_{f}d{ /\lambda )}}{1+r_{0f}r_{fs}\exp ({4\pi
}in_{f}d{/\lambda )}}\quad , \label{erefl}
\end{equation}
with $r_{0f}=(1-n_f)/(1+n_f)$ and $r_{fs}=(n_f-n_s)/(n_f+n_s)$ being the Fresnel
reflection coefficients at the air-film $(r_{0f})$ and film-substrate $(r_{fs})$
interfaces. Here $n_f = (i\sigma^* /\varepsilon_0 \omega^*)^{1/2}$ and $n_s$ are the
complex refractive indices of the film and substrate, respectively, $\lambda $ is the
radiation wavelength, $d$ is the film thickness, $\omega= 2\pi \nu$ is the angular
frequency, $\sigma^*=\sigma_1+i\sigma_2$ is the complex conductivity of the film, and
$\varepsilon_0$ is the permittivity of free space. Eq.\ (\ref{erefl}) is written
neglecting the multiple reflections from the opposite sides of the substrate. If the film
thickness is smaller than the penetration depth ($\left| n_{f}\right| d\gg \lambda $) and
if $\left| n_{f}\right| \gg \left| n_{s}\right| $, Eq. (\ref{erefl}) can be simplified to
:
\begin{equation}
r \approx \frac{1-\sigma ^{*}dZ_{0}-n_{s}}{1+\sigma ^{*}dZ_{0}+n_{s}} \label{erefl1} ,
\end{equation}
where $Z_{0}=\sqrt{\mu_0/\varepsilon_0} \simeq 377\,\Omega$ is the impedance of free
space.

\begin{figure}[]
\centering
\includegraphics[angle=270,width=8cm,clip]{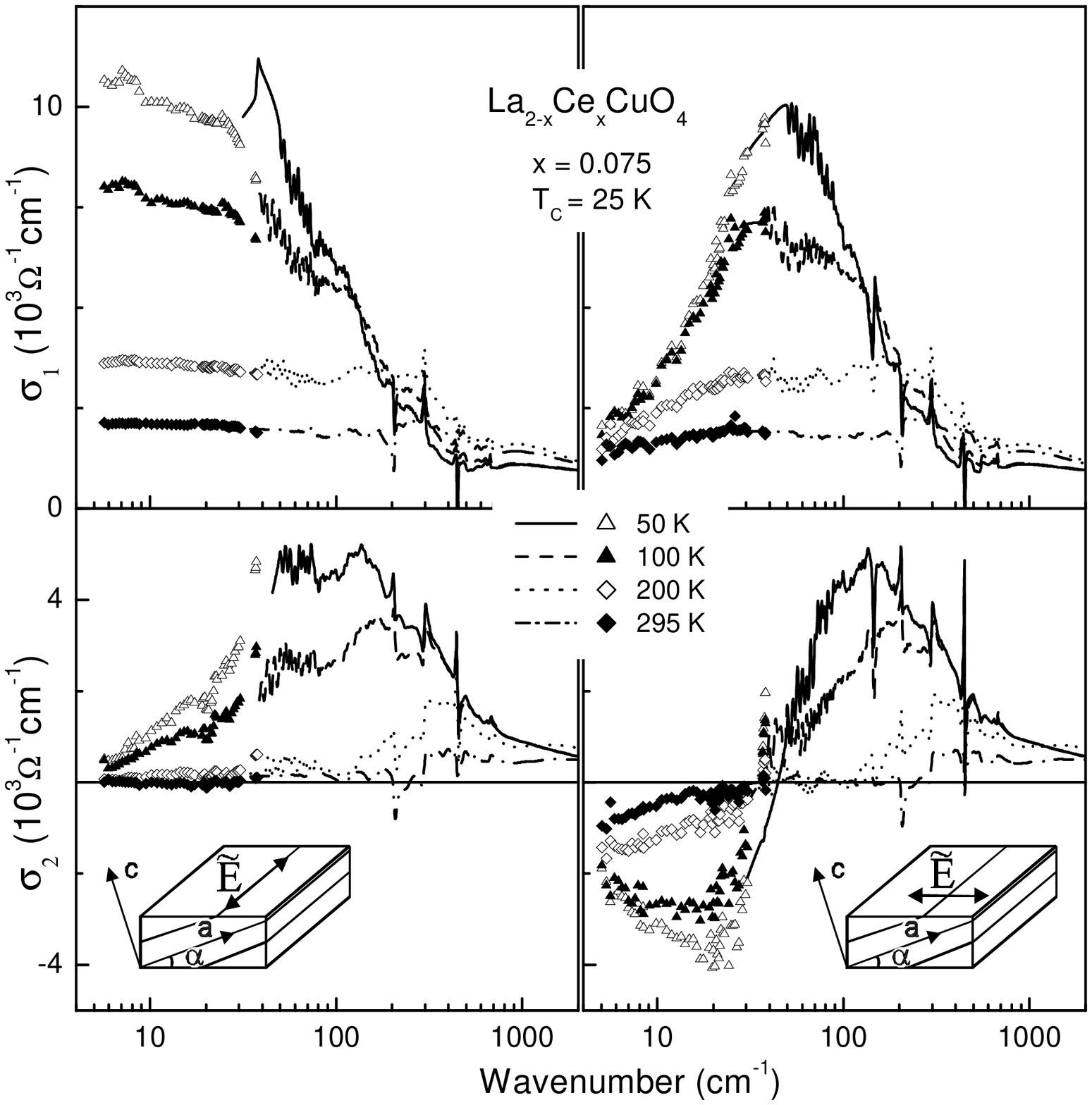}
\caption{Far-infrared conductivity of LCCO film for two different experimental geometries
as indicated in the insets. Upper panels - $\sigma_1$, lower panels - $\sigma_2$. Left
panels: in-plane response, right panels: mixed ac-geometry. Lines represent the
conductivity obtained from the infrared reflectance, symbols - the conductivity as
measured directly by the submillimeter transmission technique. } \label{fsig}
\end{figure}

Figure \ref{fsig} shows the far-infrared conductivity of LCCO. The results above
40\,cm$^{-1}$ were obtained applying the Kramers-Kronig analysis to the reflectivity data
and  solving Eq. (\ref{erefl}). Below 40\,cm$^{-1}$ the complex conductivity was
calculated directly from the transmittance and phase shift. The far-infrared conductivity
of LCCO for the in-plane response (left panels) resembles well the predictions of the
Drude model. At low frequencies, $\sigma_1$ is frequency independent and $\sigma_2$
increases approximately linearly with the frequency. For frequencies close to the value
of the scattering rate, $\sigma_1$ starts to decrease and $\sigma_2$ shows a maximum,
$\nu_{max}\simeq 1/ 2 \pi\tau$. On the contrary, the $1^\circ$ tilted geometry (right
panels) reveals dramatic changes compared to the in-plane response. The real part of the
conductivity shows an intensive peak in the far-infrared range. As expected by the
causality principle (via the Kramers-Kronig relations) the imaginary part of the
conductivity for the tilted geometry shows a step-like structure at the same frequency.
Fig. \ref{fsig} demonstrates that already $1^\circ$ off-axis geometry significantly
modifies the low-frequency conductivity spectra. However, we note that the most
significant changes appear in the low-frequency range, $\nu < 100\,$cm$^{-1}$. The
high-frequency response of the tilted geometry approximately follows the spectra of the
CuO$_2$ planes.

In order to understand the off-axis response we utilize a simple model which takes into
account a tilted geometry of the experiment \cite{ncco1,ncco2}. The model has been
developed in the assumption that the penetration  depth is mich smaller than the film
thickness. This approximation for the effective conductivity is fulfilled in the present
case and contains the essential physics of the problem. The conductivity of a thin film
tilt by an angle $\alpha$ reads:

\begin{equation}
\sigma _{eff}=\frac{-i\varepsilon _{0}\omega (\sigma _{a}\cos ^{2}\alpha +\sigma _{c}\sin
^{2}\alpha )+\sigma _{a}\sigma _{c}}{-i\varepsilon
_{0}\omega +\sigma _{a}\sin ^{2}\alpha +\sigma _{c}\cos ^{2}\alpha }\quad .%
\text{ }  \label{eqsig}
\end{equation}
Here $\sigma _{a} $ and $ \sigma _{c}$ are the complex conductivities in the
CuO$_2$-planes and along the c-axis, respectively. Within the approximation $\alpha
\approx \sin \alpha \ll 1$ and $|\sigma _{a}|\gg |\sigma _{c}|$, Eq. (\ref{eqsig}) can be
rewritten as:
\begin{equation}
\frac{1}{\sigma _{eff}}=\frac{\alpha ^{2}}{\sigma _{c}-i\varepsilon _{0}\omega}+%
\frac{1}{\sigma _{a}} \quad .
 \label{eqsimpl}
\end{equation}

To understand the spectra in Fig. \ref{fsig} qualitatively, the c-axis response of LCCO
can be taken as purely insulating with a dielectric constant $\varepsilon_c \simeq 10$.
In that case $\sigma_c \simeq -i\varepsilon_c \varepsilon_0 \omega$ and the first term in
Eq. \ref{eqsimpl} dominates the response for the tilted geometry below $\omega/2\pi
\simeq \sigma_a \alpha^2 /(\varepsilon_c \varepsilon_0)\simeq 20\,$cm$^{-1}$, which
roughly corresponds to the conductivity maximum in the right upper panel of Fig.
\ref{fsig}.

\begin{figure}[]
\centering
\includegraphics[width=7cm,clip]{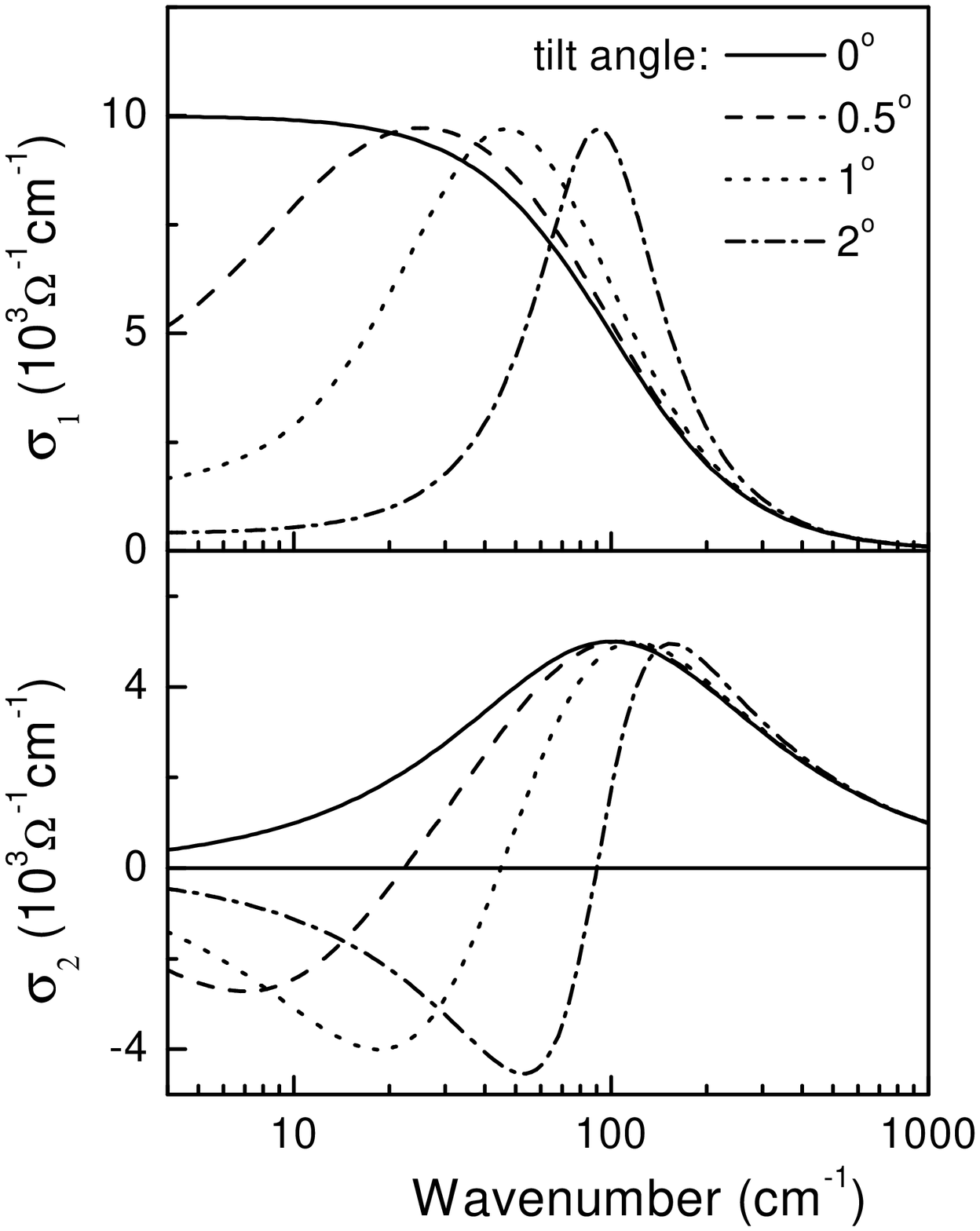}
\caption{Far-infrared conductivity of a tilt anisotropic sample as calculated from Eq.
\ref{eqsig}. The in-plane conductivity was assumed to obey the Drude-form with $\sigma_0
= 10^4\,\Omega^{-1}$cm$^{-1}$, $1/2\pi\tau = 100\,$cm$^{-1}$, and the c-axis response was
taken as $\sigma_{1,c}=0.5\,\Omega^{-1}$cm$^{-1}$ and $\varepsilon_{1,c}=8$.}
\label{fmod}
\end{figure}

Figure \ref{fmod} shows the conductivity of a tilted sample as calculated on the basis of
Eq. \ref{eqsig} and for different tilt angles $\alpha$. The parameters of the model have
been taken to correspond to the $T=50\,$K data in Fig. \ref{fsig}. The in-plane
conductivity was assumed to obey the Drude-form $\sigma^* = \sigma_0 /(1-i \omega \tau)$
with $\sigma_0 = 10^4\,\Omega^{-1}$cm$^{-1}$, $1/2\pi\tau = 100\,$cm$^{-1}$, and the
c-axis response was taken as $\sigma_{1,c}=0.5\,\Omega^{-1}$cm$^{-1}$ and
$\varepsilon_{1,c}=8$ (i.e.
$\sigma_c=\sigma_{1,c}-i\omega\varepsilon_0\varepsilon_{1,c}$). Evidently, the solid
curve ($\alpha=0$) represents the Drude response and closely resembles the in-plane
conductivity of Fig. \ref{fsig} (left panels, $T=50\,$K). Already for small tilt angle
the effective conductivity reveals a pronounced peak in the real part of the
conductivity, which for $\alpha=1^\circ$ shows the same shape as the spectra in the right
panels of Fig. \ref{fsig}.

\begin{figure}[b]
\centering
\includegraphics[width=7cm,clip]{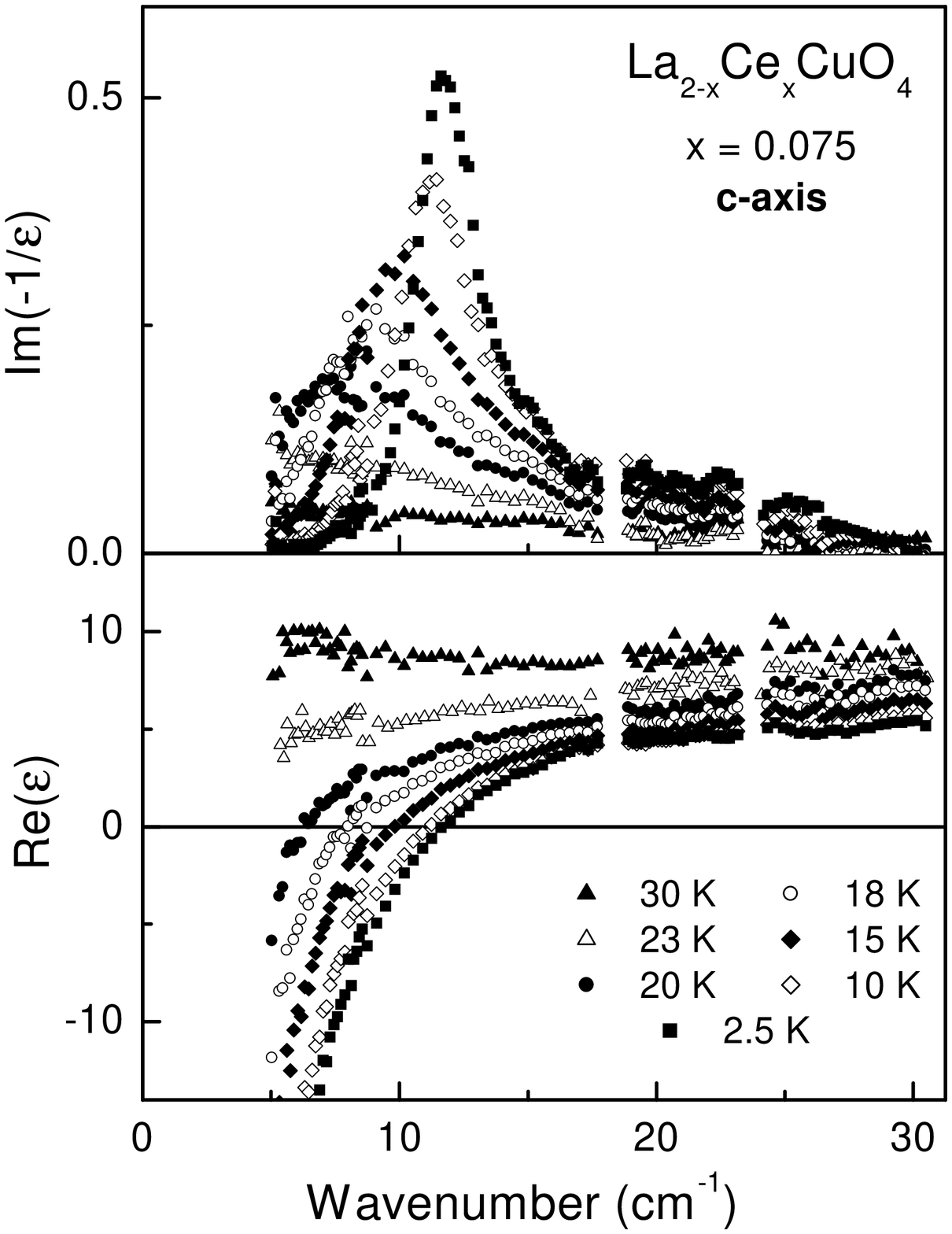}
\caption{Dielectric loss function and dielectric permittivity of LCCO along the c-axis.
The peak in the loss function corresponds to the Josephson plasma resonance.} \label{fc}
\end{figure}

Comparative analysis of the in-plane and the mixed response allows to extract the c-axis
conductivity of LCCO inverting Eq. (\ref{eqsig}). Because the spectra for both
orientations show only weak deviations at high frequencies, this procedure reveals
reliable data below $\nu \sim 30\,$cm$^{-1}$ only. The c-axis response of LCCO, obtained
in this way, is shown in Fig. \ref{fc}. The dielectric permittivity  above $T_{\rm c}$ is
approximately frequency and temperature independent with $\varepsilon_1 \simeq 8$ and
$\sigma_1 \simeq 0.5\,\Omega^{-1}$cm$^{-1}$. An additional contribution to
$\varepsilon_1$ due to the superconducting condensate, which is proportional to $-c^2
/(\omega \lambda)^2$, starts to grow below $T_{\rm c}$. Here $\lambda$ is the
temperature-dependent c-axis penetration depth, which attains the value of
$\lambda=48\,\mu$ for $T\rightarrow 0$. Correspondingly, a zero crossing of the
dielectric constant is observed in the submillimeter frequency range, which represents a
Josephson plasma resonance. This resonance is most clearly seen in the
$Im(-1/\varepsilon^*)$ presentation as a sharp maximum close to $\nu \simeq
11.7\,$cm$^{-1}$.

In conclusion, we presented the submillimeter-wave and far-infrared conductivity of a
La$_{2-x}$Ce$_{x}$CuO$_{4}$ film measured in a tilted geometry. A broad maximum in the
far-infrared conductivity can be observed within this geometry and is attributed to the
admixing of the c-axis response to the in-plane conductivity. The same effects can be
expected for strongly anisotropic metals, where highly conducting response along one
optical axis coexists with almost insulating behavior for another direction. In addition,
from the analysis of the mixed and the in-plane contributions the pure c-axis properties
of LCCO can be extracted from the spectra, which show a Josephson plasma resonance in the
superconducting state.

This work was supported by BMBF (13N6917/0 - EKM).

\end{document}